# Spin-Phonon Coupling in Iron Pnictide Superconductors

T. Egami,[1,2,3*] B. V. Fine,[4] D. J. Singh,[3] D. Parshall,[1] C. de la Cruz[1] and P. Dai[1,3]

[1]Department of Physics and Astronomy, University of Tennessee, Knoxville, TN 37996, USA,
[2] Department of Materials Science and Engineering, University of Tennessee, Knoxville, TN 37996, USA
[3]Oak Ridge National Laboratory, Oak Ridge, TN 37853, USA
[4]Institute for Theoretical Physics, University of Heidelberg, Heidelberg, 69120, Germany

*Tel.: +1 865 974 7204; fax: +1 865 974 1010. *E-mail address*: egami@utk.edu

**Abstract**

The magnetic moment in the parent phase of the iron-pnictide superconductors varies with composition even when the nominal charge of iron is unchanged. We propose the spin-lattice coupling due to the magneto-volume effect as the primary origin of this effect, and formulate a Landau theory to describe the dependence of the moment to the Fe-As layer separation. We then compare the superconductive critical temperature of doped iron pnictides to the local moment predicted by the theory, and suggest that the spin-phonon coupling may play a role in the superconductivity of this compound.

## 1. Introduction

The discovery of superconductivity in iron-pnictides [1] provoked strong interests, partly because of its similarity to the superconductivity in the cuprates. Just as in the cuprates superconductivity appears when the antiferromagnetic order in the parent phase is suppressed by doping [2], and the spin resonance peak is found by neutron scattering [3-6]. On the other hand, unlike the cuprates, the x-ray core spectroscopy shows no satellites indicating strong on-site electron correlation as in the Hubbard model [7], and the parent phase is semi-metallic [1], rather than a Mott-Hubbard insulator. A prominent property of the iron-pnictides is the strong coupling of the magnetic moment on the structure, first identified by the LDA calculations [8-10]. For the cuprates such effects have been only speculated by a model calculation [11]. We first describe this coupling for a series of compounds CeFeAs$_{1-x}$P$_x$O [12], and discuss its possible effects on the superconductivity of doped pnictides.

## 2. Landau theory of the magneto-volume effect

The parent compound CeFeAsO has a relatively large magnetic moment that orders antiferromagnetically at 140 K [13]. When As is replaced by isovalent P (CeFeAs$_{1-x}$P$_x$O) the moment decreases rapidly, and disappears at $x = 0.4$ [12]. Replacing As with P changes the lattice, most notably the distance between the Fe layer and the As/P layer ($d_{\text{As-Fe}}$). The magnitude of the moment is strongly related to $d_{\text{As-Fe}}$ as shown in Fig. 1. A possible origin of this dependence is the magneto-volume effect [14]. Because the Pauli exclusion principle operates for parallel spins the electron kinetic energy of the spin-polarized state is higher, and volume expansion relaxes the kinetic energy. Consequently the magnetic (high-spin) state has a larger volume than the non-magnetic (low-spin) state. In some iron alloys the thermal volume expansion due to lattice anharmonicity cancels the decrease in volume associated in the decrease in spin-polarization, resulting in zero thermal expansion, widely known as the Invar behavior.

This spin-lattice coupling can be described by the Landau-type theory [15]. We may write the magnetic free energy as

$$F_M = AM^2 + BM^4 - \alpha(z - z_c)M^2 \tag{1}$$

where $z = d_{As\text{-}Fe}$. By minimizing (1) we obtain

$$M = \sqrt{\frac{\alpha(z - z_c)}{2B}} \tag{2}$$

Indeed the data shown in Fig. 1 follow this dependence on $z$ quite well, and a fit to the data (solid curve) yields the parameters, $z_c = 1.278$ Å, $\alpha/2B = 13.88$ ($\mu_B$/Å). Thus the quantum critical point (QCP) at $z_c$ is characterized as the Stoner critical point [15].

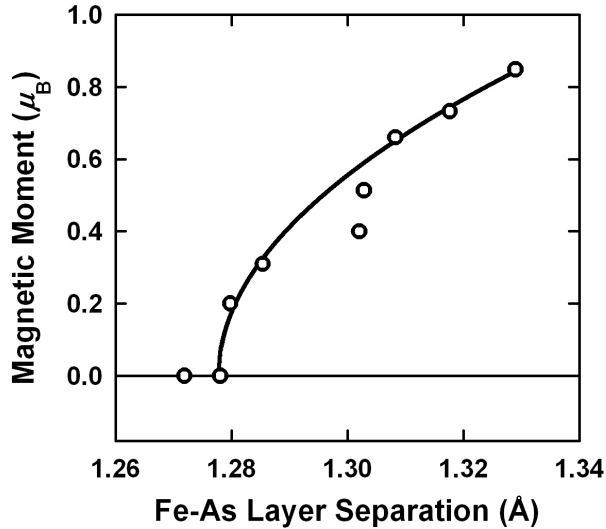

Fig. 1 Dependence of the magnetic moment on the Fe-As layer separation for CeFe(As,P)O (open circles) [12], and the fit by the Landau theory (solid curve).

## 3. Dependence of $T_C$ on structure

A remarkable dependence of the superconducting critical temperature $T_C$ on the As-Fe-As bond angle was pointed out by Lee *et al.* [16]. Because the in-plane lattice constants, *a* and *b*, do not change much with composition (± 1%), the As-Fe-As bond angle depends most sensitively on the Fe-As layer separation, $z$. Interestingly if we use eq. (2) to calculate the expected magnetic moment, we find a strong correlation between $M^2$ and $T_C$ of various iron pnictides, except for a few outliers, as shown in Fig. 2. The sources of the data for $T_C$ are found in Ref. 15. Even though the doped superconducting compositions are mostly non-magnetic, the LDA calculations suggest that their structure is compatible only with the spin-polarized state [9,17,18]. Also the core level spectroscopy suggests significant dynamic spin-polarization [6]. The present results are consistent with the dynamic local spin polarization even in the superconducting state.

The observed correlation supports the spin fluctuation mediated mechanism as the origin of superconductivity. However, it is also consistent with the electron-phonon (*e-p*) coupling through the spin-channel being responsible [15]. The strong spin-lattice coupling through the magneto-volume effect as shown by equation (2) and that fact that superconductivity occurs in the vicinity of the Stoner QCP support the possibility that this unconventional *e-p* coupling may be playing some role in the superconductivity in iron pnictides.

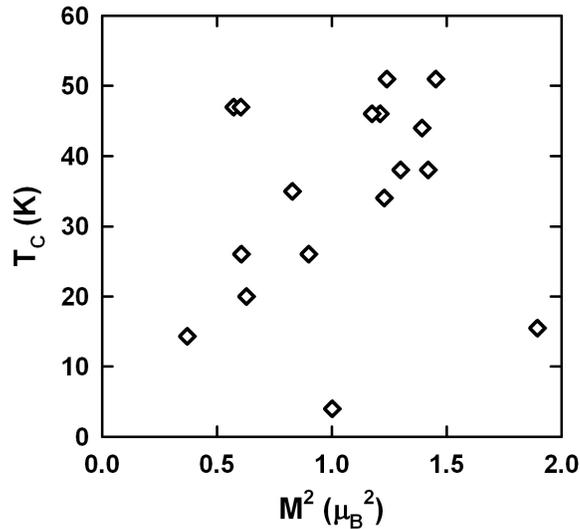

Fig. 2  Correlation between the superconducting transition temperature $T_C$ and square of the moment calculated by equation (2).

**Acknowledgment**

This work was supported by the Department of Energy through the EPSCoR grant, DE-FG02-08ER46528 and the Division of Materials Science and Engineering, Office of Basic Energy Sciences.